\documentclass[a4paper]{jpconf}
\usepackage{graphicx}
\usepackage{hyperref}
\usepackage{float}
\begin{document}
\title{Investigation of deflection angle for muon energy classification in muon scattering tomography via GEANT4 simulations}

\author{A. Ilker Topuz$^{\rm 1,2}$, Madis Kiisk$^{\rm 1,3}$, Andrea Giammanco$^{\rm 2}$, M\"{a}rt M\"{a}gi$^{\rm 3}$}

\address{$^{1}$ Institute of Physics, University of Tartu, W. Ostwaldi 1, 50411, Tartu, Estonia}
\address{$^{2}$ Centre for Cosmology, Particle Physics and Phenomenology, Chemin du Cyclotron 2, B-1348 Louvain-la-Neuve, Belgium}
\address{$^{3}$ GScan OU, Maealuse 2/1, 12618 Tallinn, Estonia}

\ead{ahmet.ilker.topuz@ut.ee, ahmet.topuz@uclouvain.be}

\begin{abstract}
In muon scattering tomography, the investigated materials are discriminated according to the scattering angle that mainly depends on the atomic number, the density, and the thickness of the medium at a given energy value. The scattering angles at different initial energies also provide the opportunity to classify the incoming muons into a number of energy groups. In this study, by employing the GEANT4 code, we show that the deflection angle exponentially decays as a function of energy, and the numerical values for the current configuration are below the detector accuracy except the initial energy bins owing to the low-Z, low density, and low thickness of the current plastic scintillators. This implies the necessity of additional components that provoke the muon scattering. Therefore, we introduce stainless steel surfaces into the top and bottom sections in order to amplify the deflection angle as well as to reduce the uncertainty, thereby improving the detector performance.
\end{abstract}
\section{Introduction}
The principle behind the muon scattering tomography is to track the propagation of the cosmic ray muons within the target volume through which the incoming muons of a certain energy deviate from their initial trajectories after a series of physical processes predominantly depending on the atomic number, the material density, and the material thickness~\cite{1}. In this study, we investigate the muon deflection due to the plastic scintillators present in our current tomographic prototype~\cite{2,3} that includes three detector layers with a thickness of 0.4 cm as well as an accuracy of 1 mrad in both the top section and the bottom section by determining the variation of the deflection angle with respect to the muon energy. We perform the Monte Carlo simulations by using the GEANT4 code~\cite{4} in order to obtain the deflection angles and we follow an experimentally replicable procedure based on the hit locations in the detector layers. 
\section{Average deflection angle and standard deviation}
\label{Average deflection angle and its standard deviation}
\begin{figure}[h]
\setlength{\belowcaptionskip}{-4ex} 
\begin{center}
\includegraphics[width=9.5cm]{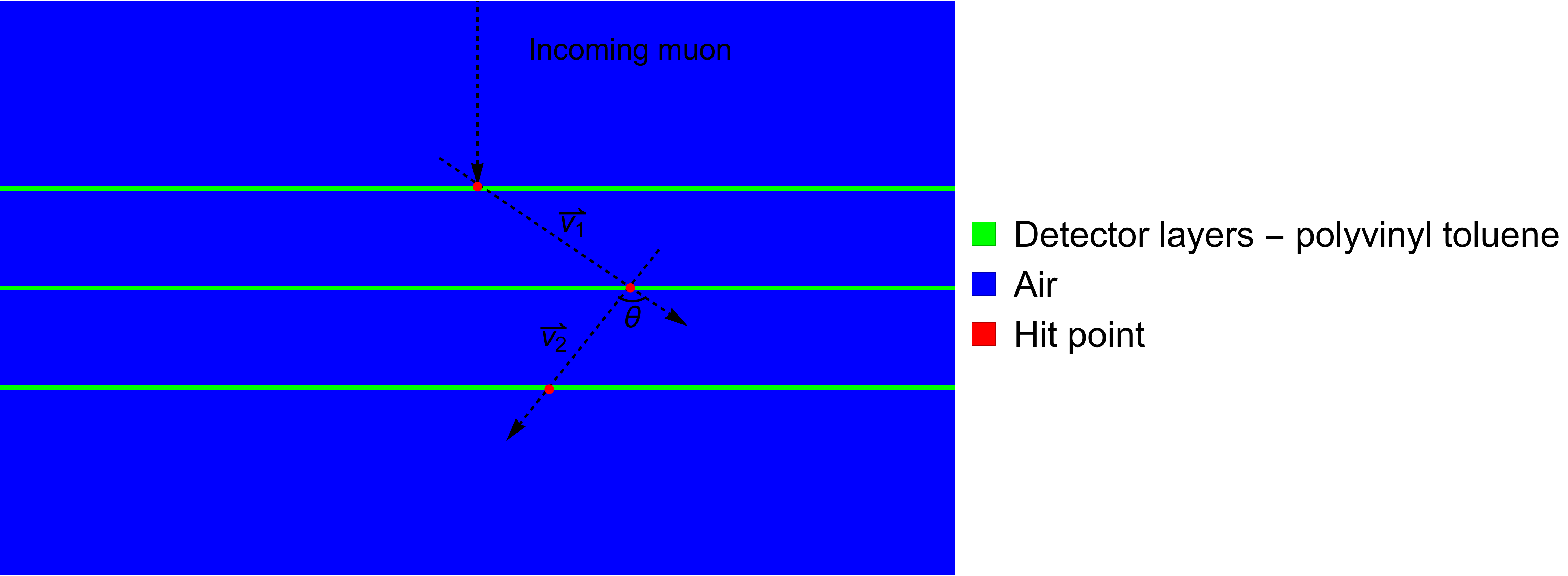}
\caption{Definition of the deflection angle denoted by $\theta$ according to the hit points in the detector layers.}
\label{deflection}
\end{center}
\end{figure}

As described in Fig.~\ref{deflection}, the computation of the deflection angle requires the construction of two separate vectors by utilizing at least three muon hit locations in the detector layers where the first vector is the difference between the second hit location and the first hit location, while the subtraction of the second hit location from the third hit location yields the latter vector.
The deflection angle is obtained by using these two vectors as follows~\cite{5}
\begin{equation}
\theta=\arccos\left (\frac{\vec{v}_{1} \cdot \vec{v}_{2}}{\left|v_{1}\right|\left|v_{2}\right|}\right)
\end{equation}
Since a significant number of muons reach the detector layers, the average profile of the deflection angle at a certain energy is calculated by averaging the previously determined deflection angles over $N$ number of the non-absorbed/non-decayed muons as written in
\begin{equation}
\bar{\theta}=\frac{1}{N}\sum_{i=1}^{N}\theta_{i}
\end{equation}
where its standard deviation is 
\begin{equation}
\delta\theta=\sqrt{\frac{1}{N}\sum_{i=1}^{N}(\theta_{i}-\bar{\theta})^{2}}
\end{equation}
In view of the fact that the deflection angle is an outcome of a stochastic process, the standard deviation of the deflection angle is expected to be reduced in order to have a better energy estimation. In essence, the average deflection angle of two different hodoscopes indicated by $x$ and $y$ at a given energy value over $N$ number of the non-absorbed/non-decayed muons yields the following expression
\begin{equation}
\bar{\theta}_{\frac{x+y}{2}}=\frac{1}{N}\sum_{i=1}^{N}\frac{\theta_{x, i}+\theta_{y, i}}{2}
\end{equation}
Consequently, its standard deviation parameterized in terms of the contributions from both the top section and the bottom section is
\begin{equation}
\delta\theta=\sqrt{\frac{1}{N}\sum_{i=1}^{N} \biggl(\frac{\theta_{x, i}+\theta_{y, i}}{2}-\bar{\theta}_{\frac{x+y}{2}}\biggr)^{2}}
\label{standarddeviationofaverageofaverage}
\end{equation}
The resulting deflection angle for our present prototype is anticipated to be very small~\cite{6}, i.e. either below or in the close neighborhood of our detector accuracy that is 1 mrad. Accordingly, the necessity of a stronger deflecting medium is foreseen in order to augment the angular deviation, and we introduce a stainless steel layer of 0.4 cm, the density of which is 8 $\rm g/cm^{3}$, into the top section and the bottom section in order to join this aim. For the sake of comparison, we also explore the coefficient of variation (CV) attributed to these two configurations with respect to the energy increase, which is defined as the ratio between the standard deviation and the average value as expressed in 
\begin{equation}
\rm CV=\frac{\mbox{Standard deviation}}{Average}=\frac{\delta\theta}{\bar{\theta}}
\label{coefficientofvariation}
\end{equation} 
\section{Hodoscope schemes and simulation properties}
\label{Hodoscope schemes and simulation properties}
\begin{figure}[h]
\setlength{\belowcaptionskip}{-4ex} 
\begin{center}
\includegraphics[width=4.8cm]{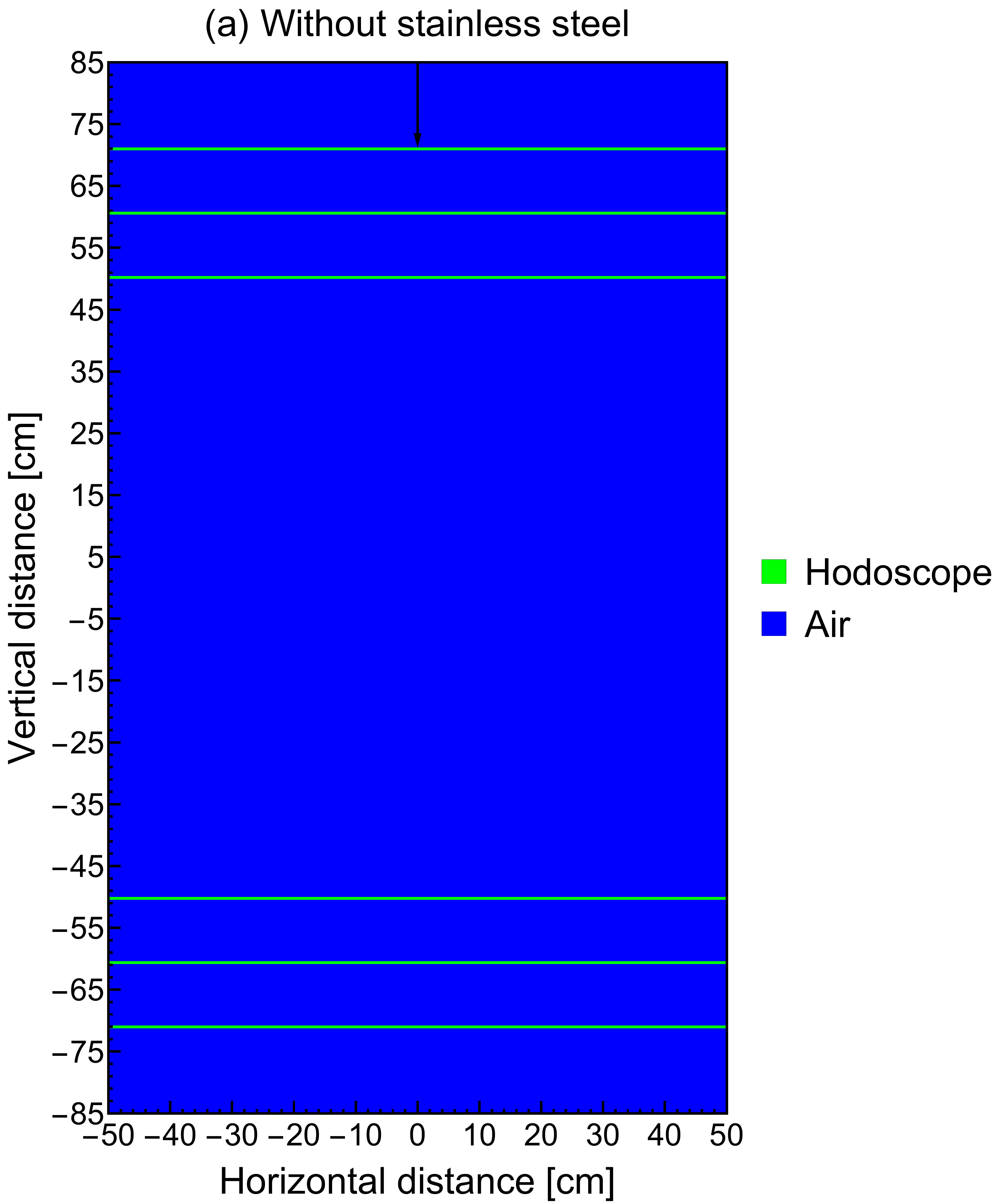}
\hskip 1cm
\includegraphics[width=5.1cm]{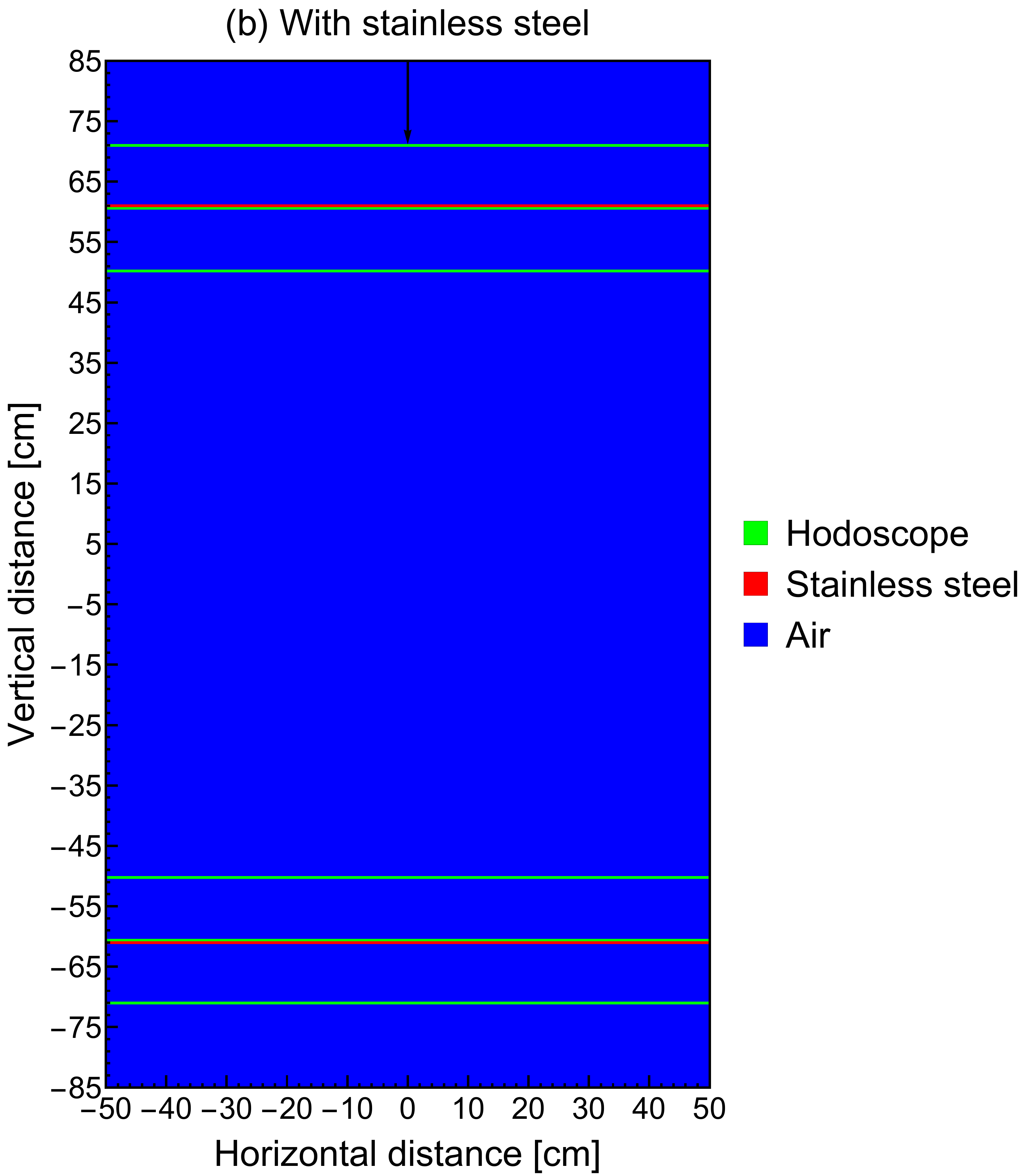}
\caption{Layouts of (a) the current hodoscope without the stainless steel layers (b) the alternative hodoscope with the stainless steel layers.}
\label{Without and with stainless steel}
\end{center}
\end{figure}

The geometrical schemes for either setup are depicted in Figs.~\ref{Without and with stainless steel}(a) and (b), and it is seen that the detector layers are separated by a distance of 10 cm, whereas the span between these two hodoscopes is 100 cm. Furthermore, the dimensions of both the detector layers and the stainless steel layers are $100\times0.4\times100$ $\rm cm^{3}$.
We use a central mono-energetic mono-directional beam that is generated at y=85 cm via G4ParticleGun, and the generated muons are propagating in the vertically downward direction, i.e. from the top edge of the simulation box through the bottom edge. Noting that the distribution of the incident angle ($\alpha$) is approximated via $\cos^2(\alpha)$ for an interval between $-\pi/2$ and $\pi/2$~\cite{7}, this source setup stands for a feasible approach since the present aperture of the entire detection geometry typically only covers the narrow angles besides the very rare entries around the corners. At a given energy, the number of the simulated muons is 1000, and we investigate the deflection angle for an energy interval between 0.5 and 8 GeV with an increase step of 0.5 GeV. All the materials in the current study are defined according to the GEANT4/NIST material database. The reference physics list used in these simulations is FTFP$\_$BERT.

\section{Comparison between hodoscopes without/with stainless steel layers}
\begin{figure}[h]
\begin{center}
\setlength{\belowcaptionskip}{-4ex} 
\includegraphics[width=6.5cm]{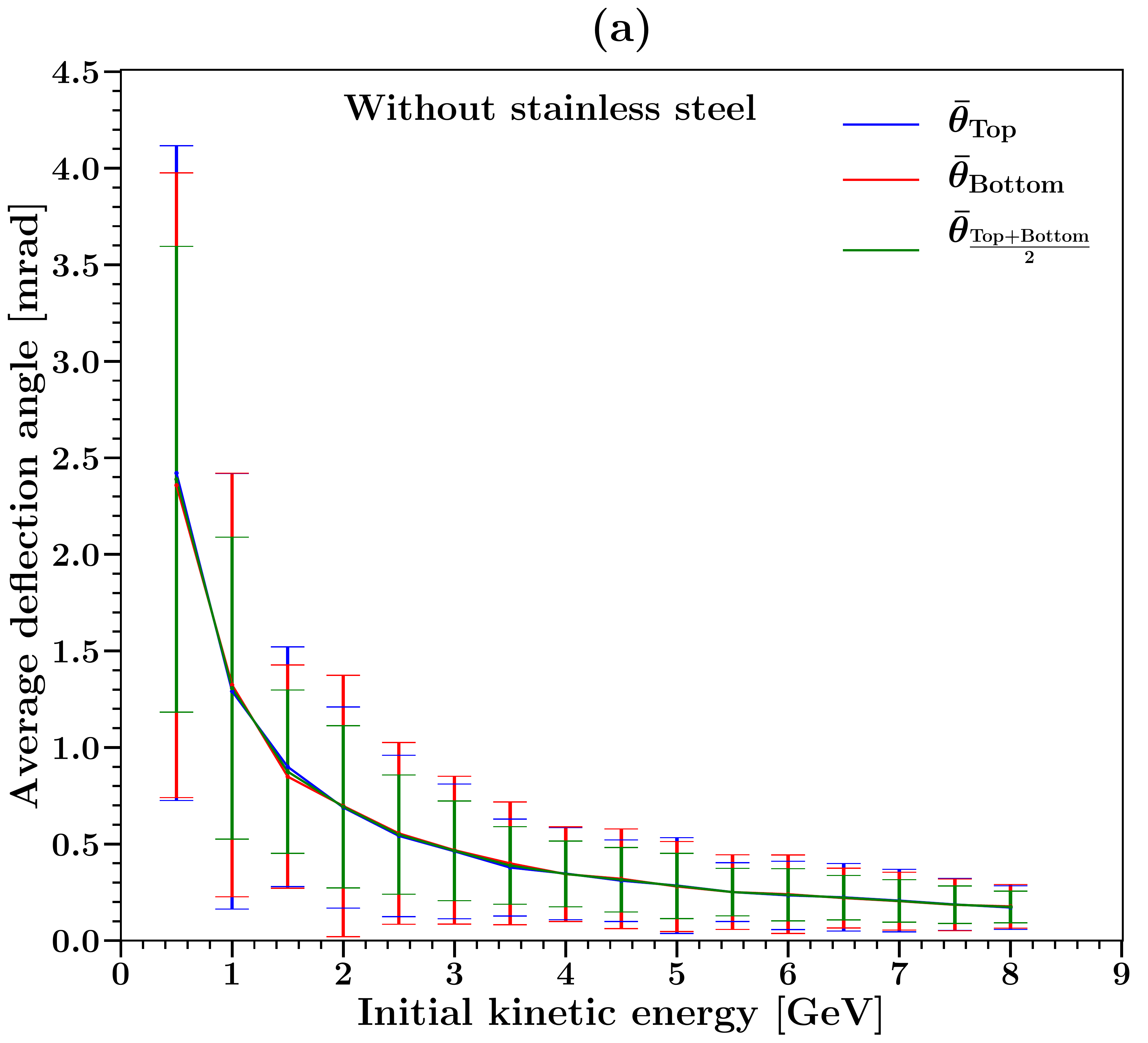}
\includegraphics[width=6.4cm]{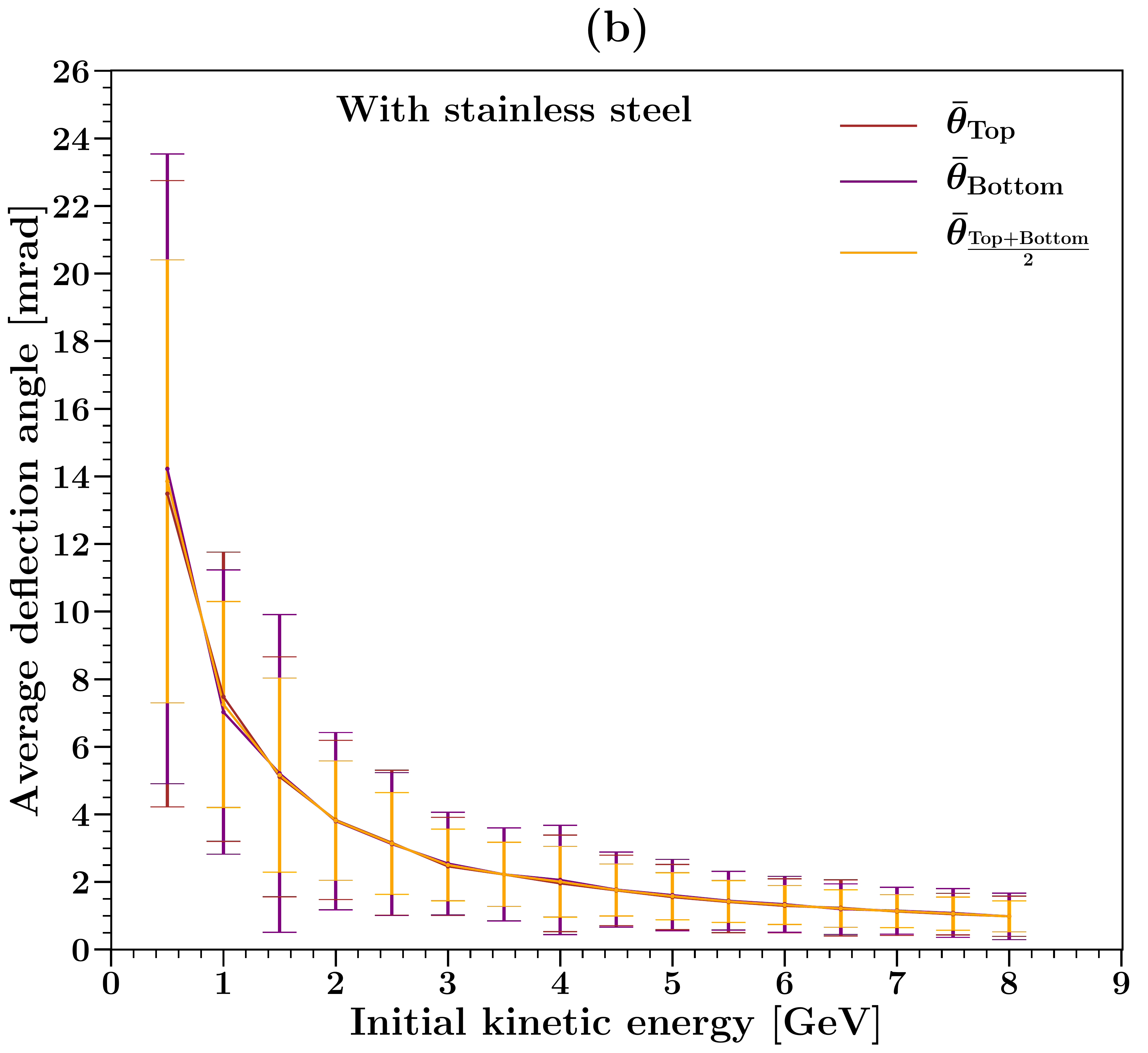}
\caption{Variation of the average deflection angles versus the energy increase (a) in the present hodoscope (b) in the alternative hodoscope with the stainless steel layers.}
\label{Hodoscopewithoutwithsteel} 
\end{center}
\end{figure}

\begin{figure}[h]
\begin{center}
\setlength{\belowcaptionskip}{-4ex} 
\includegraphics[width=6.2cm]{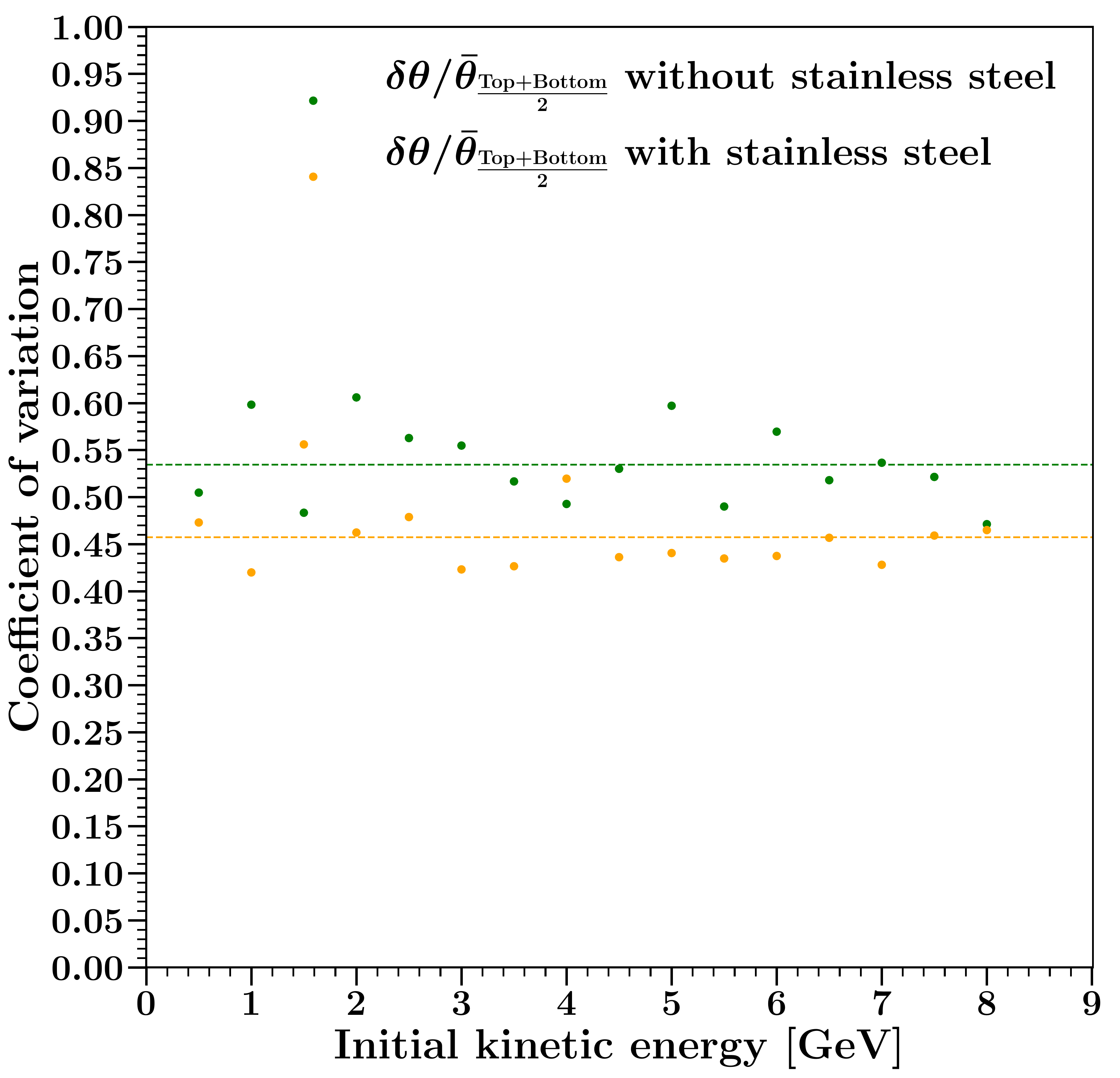}
\caption{Effect of the stainless steel layers on the trend in the coefficient of variation.} 
\label{coefficientofvariations}
\end{center}
\end{figure}

We initiate our investigation based on our first configuration by showing the simulation results for the average deflection angles in Fig.~\ref{Hodoscopewithoutwithsteel}(a). In connection with the energy increase, the observed trend in the mean deflection angle is the exponential decay. Due to the exact symmetry in the structural composition in both the top section and the bottom section, the simulation outcomes are similar in either section. It is also seen that the resulting deflection angles are either under the detector accuracy of 1 mrad or in its close periphery, and this is a principal problem to be addressed in the present study. Regarding the standard deviations, it is demonstrated that the deflection angles are dispersed widely, rather than converging in a narrow interval, which sets the energy categorization more challenging. As described in Eq.~(\ref{standarddeviationofaverageofaverage}), the consecutive averaging sequentially over the number of sections and the number of the non-absorbed/non-decayed muons leads to a significant reduction in the angular width.
After the qualitative examination of the average deflection angles for our current hodoscope, an experimental limitation that awaits to be attacked is our present detector accuracy of 1 mrad. As an alternative to a significant detector upgrade that captures the angular values below 1 mrad, we propose the introduction of stainless steel layers with an optimally low thickness of 0.4 cm as well as a density of 8 $\rm g/cm^{3}$, which arouse the angular deflection, into either section as depicted in Fig.~\ref{Without and with stainless steel}(b). We repeat our simulations on this alternative configuration by using the same simulation features, and the average deflection angles along with the standard deviations for the new tomographic setup with the stainless steel layers are displayed in Fig.~\ref{Hodoscopewithoutwithsteel}(b). In addition to the notable increase in the average deflection angles, the angular width is also remarkably reduced except the fluctuations at a couple of energy values, i.e. 1.5 and 4 GeV. Concerning the visible influence of the stainless steel layers on the standard deviations, Fig.~\ref{coefficientofvariations} shows the trends in coefficients of variation as expressed in Eq. (\ref{coefficientofvariation}) with the intention of comparing both detector setups, and it is concretely seen that the angular uncertainty is lowered by the presence of the inserted stainless steel layers, which also means that an improved energy classification is expected from the proposed alternative scheme.
\section{Conclusion}
We ameliorate the width of the deflection angle by averaging the simulation outcomes from the top detector layers and the bottom detector layers over the number of the non-absorbed/non-decayed muons. Since the average deflection angles are mostly below the current detector accuracy that is 1 mrad, we introduce the stainless steel layers to augment the average deflection angles as well as to further diminish the standard deviations.
\label{Conclusion}

\end{document}